\documentclass[%
 reprint,
superscriptaddress,
nofootinbib,
nobibnotes,
bibnotes,
 amsmath,amssymb,
 aps,
]{revtex4-2}

\usepackage{graphicx}
\usepackage{dcolumn}
\usepackage{bm}
\usepackage{amscd}
\usepackage{caption}
\usepackage{subcaption}
\usepackage{float}
\usepackage{siunitx}
\usepackage{natbib}
\usepackage{hyperref}

\DeclareMathOperator{\Var}{\widehat{Var}}
\begin{document}

\preprint{APS/123-QED}

\title{Image Classification with Rotation-Invariant Variational Quantum Circuits}

\author{Paul San Sebastian Sein}
\affiliation{
Ikerlan Technology Research Centre, Basque Research and Technology Alliance (BRTA),
Paseo J.M. Arizmediarrieta 2, E-20500, Arrasate-Mondragón, Spain}%
\affiliation{
Universidad del País Vasco/Euskal Herriko Unibertsitatea UPV/EHU}%

\author{Mikel Cañizo}

\affiliation{
Ikerlan Technology Research Centre, Basque Research and Technology Alliance (BRTA),
Paseo J.M. Arizmediarrieta 2, E-20500, Arrasate-Mondragón, Spain}

\author{Román Orús}
\affiliation{Donostia International Physics Center, Paseo Manuel de Lardizabal 4, E-20018 San Sebasti\'an, Spain}
\affiliation{Multiverse Computing, Paseo de Miram\'on 170, E-20014 San Sebasti\'an, Spain}
\affiliation{Ikerbasque Foundation for Science, Maria Diaz de Haro 3, E-48013 Bilbao, Spain}

\begin{abstract}
Variational quantum algorithms are gaining attention as an early application of Noisy Intermediate-Scale Quantum (NISQ) devices. One of the main problems of variational methods lies in the phenomenon of \textit{Barren Plateaus}, present in the optimization of variational parameters. Adding geometric inductive bias to the quantum models has been proposed as a potential solution to mitigate this problem, leading to a new field called Geometric Quantum Machine Learning. In this work, an equivariant architecture for variational quantum classifiers is introduced to create a label-invariant model for image classification with $C_4$ rotational label symmetry. The equivariant circuit is benchmarked against two different architectures, and it is experimentally observed that the geometric approach boosts the model's performance. Finally, a classical equivariant convolution operation is proposed to extend the quantum model for the processing of larger images, employing the resources available in NISQ devices.
\end{abstract}

\maketitle


\section{Introduction}\label{sec:intro}

Together with the recent surge of quantum technologies, both in theoretical frameworks and practical applications, quantum computing has been placed in the spotlight. Many quantum algorithms have shown significant speed-ups with respect to classical algorithms and a wide variety of applications are being explored, from simulation of many-body systems to combinatorial optimization, cryptanalysis, database search, machine learning, finance, and beyond \cite{fedorov_quantum_2022, abbas_quantum_2023,au-yeung_quantum_2023, qcfin, qccyber, qcopt1, qcopt2, qcopt3}. When talking about \emph{speed-ups} or \emph{quantum advantage}, it is referred to proven algorithmic improvement of the computational complexity needed to solve a certain problem, compared to the best known classical algorithm \cite{shorPolynomialTimeAlgorithmsPrime1997, groverFastQuantumMechanical1996,  deutschRapidSolutionProblems1997}. However, many of these algorithms require hardware resources that stand far from the capabilities of the currently available noisy intermediate-scale quantum (NISQ) devices. For this reason, and in parallel to foundational algorithmic research focused on a quantum advantage in fault-tolerant quantum devices, a significant effort is put into the search for mid-term applicability and utility of algorithms in the NISQ era. Although its computational advantage is yet to be proven, Quantum Machine Learning (QML) seems one of the most promising candidates for an early application of quantum devices \cite{wittek_quantum_2014, biamonte_quantum_2017, schuld_machine_2021}, since in many algorithms the goal is to approximate a distribution and the result often does not rely on an exact value of the computation. This fact makes QML algorithms resilient to noise, and some signs hint that the noise could even improve the optimization in specific cases \cite{liu_stochastic_2023, wang_comprehensive_2024}.

QML encompasses a broad meaning, but the scope of this work will focus on the employment of quantum circuits in learning problems that involve classical data. Variational quantum algorithms \cite{cerezo_variational_2021,havlicek_supervised_2019,farhi_classification_2018, rodriguez-grasa_training_2024} hold a renowned place in this paradigm, where quantum circuits are used to create expressive high-dimensional feature maps that have been proved to be asymptotically universal approximators \cite{perez-salinas_data_2019, schuld_effect_2020}. However, these variational algorithms suffer trainability issues called \textit{Barren Plateaus} (BP), where the variance of the loss function tends to vanish exponentially with the size of the quantum circuit \cite{mcclean_barren_2018,holmes_connecting_2022, holmesBarrenPlateausPreclude2021}. Extensive work has been done characterizing this phenomenon, identifying its sources \cite{larocca_diagnosing_2022, larocca_theory_2023, cerezo_cost_2021, ragone_unified_2023-1, fontana_adjoint_2023-1, pattiEntanglementDevisedBarren2021, ortizmarreroEntanglementInducedBarrenPlateaus2021, arrasmithEffectBarrenPlateaus2021} and looking for methods to mitigate or avoid it \cite{park_hamiltonian_2024, liu_mitigating_2023, sack_avoiding_2022}.

Geometric Quantum Machine Learning (GQML) \cite{nguyen_theory_2022,ragone_representation_2023, meyer_exploiting_2023,larocca_group-invariant_2022}, has shown a great potential to alleviate \textit{Barren Plateaus} and bring significant improvements in trainability and generalization of Variational Quantum Classifiers \cite{schatzki_theoretical_2022}. Classical Geometric Deep Learning \cite{bronstein_geometric_2021} founded an algebraic framework to analyze learning problems and proposed a geometric approach to add \textit{inductive bias} \cite{bowles_contextuality_2023} to learning models, i.e., a blueprint to design algorithms based on the prior knowledge about the geometric properties of the data set. As its quantum counterpart, GQML has established a potential research line where many authors implemented the geometric approach to a variety of practical problems, showing promising results. For example, Tüysüz, Cenk, et al. studied the behavior of equivariant quantum neural networks (EQNN) under the presence of noise and suggested a strategy to enhance the protection of symmetry \cite{tuysuz_symmetry_2024}; Zheng, Han, et al. introduced an EQNN for a classification task with $\mathbb{Z}_2$ permutation symmetry \cite{zheng_benchmarking_2022}; East, R. D., et al. built a $SU(2)$- equivariant quantum circuit to learn on spin networks \cite{east_all_2023}; Heredge, Jamie, et al. proposed a data-encoding strategy for Point Clouds, which is invariant under point permutations \cite{heredge_permutation_2023}; and Skolik, Andrea, et al. designed a permutation-equivariant quantum circuit to solve the Traveling Salesman Problem and train it using Reinforcement Learning \cite{skolik_equivariant_2022}.

All this work contributes significantly to the state-of-the-art of GQML,  but only a few articles can be found that apply the geometric approach to the problem of image classification, which is one of the main tasks in supervised learning. Among them, West, M. T., eta al. presented an equivariant quantum classifier for images with reflection label-symmetry \cite{west_reflection_2023} and a rotation equivariant Quantum Neural Network for scanning tunneling microscope images, employing Quantum Fourier transform \cite{west_provably_2023}. Additionally, Chang, S. Y., et al. proposed an equivariant Quantum Convolutional Neural Network for images with $p4m$ label-symmetry \cite{chang_approximately_2023}. Nevertheless, all these works employ the same data encoding method, namely  \textit{amplitude encoding}, which offers an exponential qubit reduction, but at the same time, a circuit depth overhead that scales exponentially with the number of qubits \cite{schuld_machine_2021, ashhabQuantumStatePreparation2022a, wangEfficientQuantumAlgorithm2009, weigoldExpandingDataEncoding2021}. On the other hand, a gap is identified in the literature, covering the application of GQML to image classification, employing \textit{angle encoding}, which allows an efficient application of re-uploading layers, boosting the approximation ability of the model arbitrarily with the number of applied layers. This encoding can be implemented in linear depth (see Sec.\ref{subsec:anglencoding}), and thus, can be executed in NISQ devices.

In this context, this paper aims to introduce an architecture for a variational quantum circuit for image classification, equivariant to $C_4$ image-rotation label-symmetry, and using angle encoding. The simple choice of the data encoding scheme determines completely the gate symmetrization of the quantum circuit, which makes the proposed architecture differ from the aforementioned related work. In this case, the $C_4$ symmetry group is considered, but more symmetry operations, such as image reflection symmetry, can be taken into account following the same methodology of gate symmetrization. However, reflection symmetry is not included in the scope, making the proposed model distinguish chiral images. This makes the model fit, for example, for medical vision problems like brain tumor image detection or fingerprint classification, where it is vital to discern between the left and right sides of the body. Additionally, since one of the bottleneck of angle encoding is the required number of qubits for the codification, equivariant classical convolution layers are proposed to reduce the dimensionality of the data, while preserving the symmetry properties of the whole model. This enables the quantum model to handle large resolution images.

Concretely, the contributions of this work are as follows:

\begin{itemize}
    \item The design of an $C_4$ invariant Variational Quantum Classifier. A detailed explanation of the followed methodology, as well as a benchmarking and a comparison with the other two architectures.
    \item A $C_4$ equivariant convolution operation to extend the label-invariant quantum model for large-scale multichannel images. 
\end{itemize}

The paper is organized as follows: in Sec.\ref{sec:preliminaries}, a formal explanation of the classification problem is given, the general scheme of  Variational Quantum Classifiers (VQC) is presented, and the intuition behind GQML is summarized. In Sec.\ref{sec:results}, the concrete application of the methodology is explained, and the benchmarking results are shown. First, in Sec.\ref{subsec:imagedata}, the characteristics of the synthetic data set generated to design the circuit are presented. Then, in Secs.\ref{subsec:anglencoding} and \ref{subsec:Equivariant variational}, the chosen encoding strategy and the design of the equivariant VQC are detailed. In Sec.\ref{subsec:simulations} the comparison between results of different models run in simulators is done, and in Sec.\ref{subsec:scalling}, a method to extend the model to process larger images is proposed. Finally, in Sec.\ref{sec:results}, some conclusions and suggestions about future work are given.

\section{Preliminaries}\label{sec:preliminaries}

\subsection{Classification problem}
Machine Learning models are meant to emulate a decision-making process that follow intelligent beings. In this process, a response $y \in \mathcal{Y}$ is computed as a function of a piece of input information $x\in \mathcal{X}$. Therefore, there exists a function, often called \textit{hypothesis}, that computes $h: \; \mathcal{X} \longrightarrow \mathcal{Y}$ mapping. A \textit{model} is an algorithm designed to give a response $\hat{f} \in \mathcal{Y}$, which does not have to coincide necessarily with $y$ in an initial stage, for an input data-point $x$. A key characteristic of these algorithms is that they contain intern parameters $\theta$ that can be tuned to modify their output for a given data point. Hence, the mapping $\hat{h}:\; \mathcal{X} \longrightarrow \mathcal{Y}$ that it computes can be modified. The main task in Machine Learning is to find the optimal configuration of $\theta$ parameters to make the model's output the most similar possible to the output of the target process for all inputs; or, in other words, to make $\hat{h}$ approximate $h$. This is what \textit{training} the model means, or what it is referred to when stating that a machine can \textit{learn}.

In supervised learning, a data set $\mathcal{D} = \{(x^{(1)},y^{(1)}),\dots,(x^{(N)},y^{(N)})\}$ is provided with $N$ sample data-points and their correspondent responses. A \textit{loss function} $\mathcal{L}$ is defined, which quantifies the dissimilarity between the hypothesis function and the approximation of the model based on the available data set. The training of the model can be translated into an optimization problem where the optimal parameters will be the ones that minimize the loss function, following the so-called \textit{empirical risk minimization} strategy\cite{vapnik_principles_1991}:
\begin{equation}
    \theta^* = arg\,\min_{\theta} \sum_i^N \mathcal{L} (y^{(i)}, \hat{y}(x^{(i)},\theta) ).
\end{equation}

In classification problems, the response is often discretized into categorical (non-numerical) \textit{labels}. The most straightforward case is the binary classification problem, where the continuous output $f(x, \theta )$ of an algorithm can be discretized, establishing a decision threshold, for example, $ \hat{y}(x,\theta) = sign(f(x,\theta))$.

\subsection{Variational Quantum Classifier}
Variational Quantum Classifiers (VQC) are hybrid algorithms, which make use of quantum machines to give a prediction out from classical data and employ classical computers to optimize the tuneable parameters. On the one hand, the quantum circuit is composed of an \textit{encoding unitary} $U(x)$ that encodes classical data features into a quantum state. On the other hand, a set of parameterized gates is applied, preceded by multi-qubit gates that generate entangling, altogether forming a \textit{variational unitary} $W(\theta)$ \cite{schuld_machine_2021,havlicek_supervised_2019}. The encoding and variational unitaries will evolve the initial state of the system into another quantum state, function of the classical input data and the variational parameters:
\begin{equation}
    |\psi(x,\theta)\rangle = W(\theta) U(x) |0\rangle^{\otimes n_q},
\end{equation}
where $n_q$ is the number of qubits in the quantum circuit. The predicted output of a data-point is defined as the expected value of an observable $O$ for the quantum state $|\psi(x,\theta)\rangle$:
\begin{eqnarray}
    f(x,\theta) & = & \langle \psi(x,\theta)|O|\psi(x,\theta)\rangle \nonumber \\
   &=& \langle 0|U^\dag(x)W^\dag(\theta)OW(\theta)U(x)|0\rangle .
\end{eqnarray}

The outputs of the quantum circuit are then evaluated in the loss function, where a classical optimizer will determine the update of variational parameters, making use of techniques such as \textit{parameter shift} \cite{wierichs_general_2022} to compute the gradient of the loss function with respect to the optimizable parameters.

\begin{figure}
    \centering
    \includegraphics[scale = 0.85]{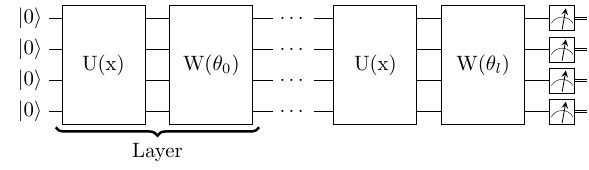}
    \caption{\label{fig:reuploading} A general scheme of a data re-uploading VQC of $4$ qubits and $l$ layers, where a layer is defined as a sequence of data encoding and variational unitaries. }
\end{figure}

Repeating the sequence of encoding and variational unitaries, widely known as \textit{data re-uploading}  \cite{perez-salinas_data_2019} (see Fig. \ref{fig:reuploading}),  has shown a significant improvement in the model's approximation capability. Furthermore, \cite{schuld_effect_2020} claim that a data re-uploading model can be considered as a Fourier series, where the accessible frequency spectrum is directly related to the number of re-uploading layers. Therefore, data re-uploading VQCs can be considered asymptotically universal approximators.

\subsection{Geometric Quantum Machine Learning}\label{subsec:GQML}

It is said that a data set contains a \textit{symmetry} if a certain property of the data set is preserved under the action of an operation, namely a \textit{symmetry operation}. The set of all such  operations $g$ along with a composition operation between them, constitute a \textit{group} $G$ \cite{davvaz_groups_2021, cornwell1997group, hamermesh2012group}. It is important to keep in mind that this symmetry group $G$ is defined in general and abstract terms, and it can act over different mathematical structures through its different \textit{representations} \cite{ragone_representation_2023}. To illustrate this, it can be thought of two representations of the same data-point linked by a data encoding unitary $U(x)$. This unitary maps a data-point from its original space $\mathcal{X}$ to the \textit{Hilbert space} $\mathcal{H}$ of our quantum circuit. The symmetry operation $g \in G$ then will have two representations acting on each one of the spaces the data-point is represented:
\begin{equation}
    \begin{CD}
        \mathcal{X} @>V_g>> \mathcal{X} \\
        @VVU(x)V @VVU(x')V\\
        \mathcal{H} @>U_g>> \mathcal{H}
    \end{CD}    
\end{equation}
where $V_g$ is the representation of $g\in G$ acting on the original data-space and $U_g$ its representation acting on the Hilbert's space. Quantum gates are represented as unitary square matrices from the group $U(2^n)$, where $n_q$ stands for the number of qubits they act on. In consequence, it is possible to represent $U_g$ as a $2^{n_q}\times 2^{n_q}$ unitary matrix, but it is remarkable how its ultimate representation will be determined by the choice of the data encoding method $U(x)$. 

The purpose of this framework is to provide some geometric tools to design \textit{label-invariant} VQCs with respect to the action of $G$:
\begin{eqnarray}
 y(x) = y(V_g[x]) & \Longrightarrow & f(x,\theta ) = f(V_g[x], \theta) \nonumber \\
  & & \forall x \in \mathcal{X} \;\textrm{and} \; \forall g\in G,
\end{eqnarray}
being $f$ the output response of the model. To achieve this goal, the first step is to ensure that the chosen encoding unitary is \textit{equivariant} with respect to the symmetry group \cite{meyer_exploiting_2023}: 
\begin{eqnarray}\label{eq:equiv_encoding}
    U(V_g[x]) = U_gU(x)U_g^\dag \nonumber\\
    \forall x \in \mathcal{X} \;\textrm{and} \; \forall g\in G.
\end{eqnarray}

If the condition of Eq. \ref{eq:equiv_encoding} is satisfied, an \textit{equivariant re-uploading model} can be built, symmetrizing all the gates used in the variational part. This is done using the \emph{Twirling formula} \cite{meyer_exploiting_2023,nguyen_theory_2022}:
\begin{equation}
    \mathcal{T}[A] = \frac{1}{|G|}\sum_{g\in G} U_g A U_g^\dag,
\end{equation}
which, applied to a unitary $A$, returns another unitary that commutes with all symmetry operations in the symmetry group $[\mathcal{T}(A), U_g]=0 \;\; \forall g \in G$. Accordingly, if the variational unitary $W$ is built out from symmterized gates, it will satisfy $[W(\theta), U_g ]=0$, and all re-uploading layers $U_l(x,\theta_l) = U(x)W_l(\theta_l)$ will be equivariant with respect to the symmetry group.

Finally, if an \textit{invariant initial-state} $|\psi_0 \rangle =  U_g |\psi_0 \rangle $ and an \textit{invariant observable} $O = U_g O U_g^\dag$ are chosen, both  for $\forall g \in G$, a label-invariant VQC will be obtained:
\begin{eqnarray}\label{eq:proof_invariance}
    f(V_g[x], \theta) &= &\langle \psi_0 |U^\dag(V[x])W^\dag(\theta)OW(\theta)U(V[x])|\psi_0 \rangle \nonumber \\
     &= &\langle \psi_0 |U_g U^\dag(x)U_g^\dag W^\dag(\theta)OW(\theta)U_gU(x)U_g^\dag |\psi_0 \rangle\nonumber \\
    & = & \langle \psi (x,\theta)|U_g^\dag O U_g |\psi(x,\theta)\rangle \nonumber \\
    &=& \langle \psi (x,\theta)| O |\psi(x,\theta)\rangle\nonumber \\
    &=& f(x, \theta).
\end{eqnarray}

Recall that as $U_g$ is unitary, $U_g^\dag = U_g^{-1}$ is also a symmetry operation and $U_g^{-1}=U_{g^{-1}}$ . This property has been used to proof the label-invariance in Eq. \ref{eq:proof_invariance}.

\section{Methodology} \label{sec:results}
The objective of this work is to apply the tools explained in Sec. \ref{subsec:GQML} to an image classification problem with 4-fold rotational label-symmetry and observe experimentally the existence of an improvement in the performance of symmetry-preserving models. As it will be discussed later in the article, the number of pixels encoded in the quantum circuit is a limiting factor for the simulation of quantum circuits in classical computers. For this reason, the size of the problem has been constrained to a reduced dimension, but the obtained results can be generalized to larger image resolutions without loss of generality.

\subsection{Synthetic toy dataset}\label{subsec:imagedata}
\begin{figure}
    \begin{subfigure}[b]{0.45\columnwidth}
        \includegraphics[width=\columnwidth]{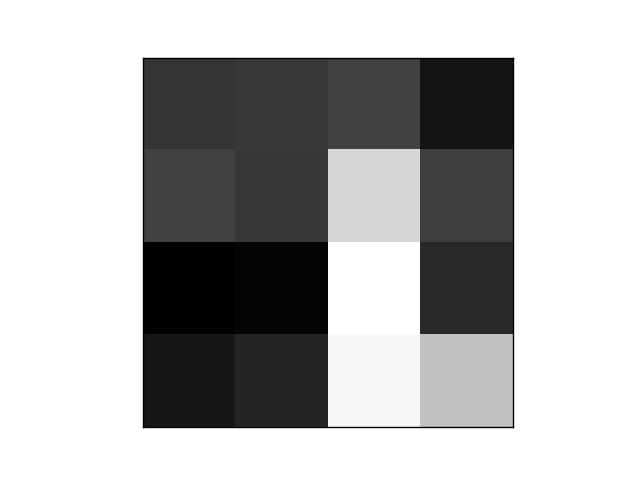}
        \caption{\textit{L} tetromino}
        \label{fig:subfigA}
    \end{subfigure}
    \hfill
    \begin{subfigure}[b]{0.45\columnwidth}
        \includegraphics[width=\columnwidth]{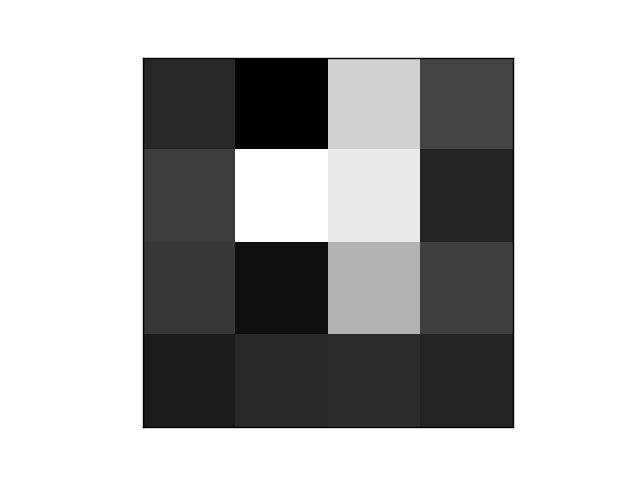}
        \caption{\textit{T} tetromino}
        \label{fig:subfigB}
    \end{subfigure}
    \caption{\label{fig:tetrominoes}Two samples of the data set.}
    \label{fig:twoSubfigures}
\end{figure}

The employed data set will be formed by square images of $n\times n$ pixels, where each pixel will take values laying between 0 and 255 in grayscale. Hence, each image can be represented by a matrix $x\in GL_n(\mathbb{R})$. For a first stage, images with resolution $n=4$ of \textit{T} and \textit{L} tetrominoes \cite{golomb1996polyominoes} have been generated (see Fig. \ref{fig:tetrominoes}). Each tetromino can be placed in 6 different positions and 4 orientations inside an image of $4\times4$ pixels, but it is possible to augment the size of the data set arbitrarily by adding a layer of noise to each image.

Tetrominoes contain a 4-fold rotation label-symmetry, but notice how reflection symmetry is not fulfilled since the chiral image of a \textit{L} tetromino is labeled as \textit{J}, breaking label invariance. A 4-fold rotation, i.e. $90^{\circ}$ rotation of the image can be expressed as a reordering of the pixels described by the  following index transformation:
\begin{equation}\label{eq:rotation_index}
    x'_{i,j} = x_{n-1-j,i}
\end{equation}

where $x_{i,j}$ is a matrix elements representing a pixels value. A crucial step to build a label-invariant quantum model is to find the representation of this operation acting on the Hilbert space of the quantum circuit. To do so, first, the chosen data-encoding strategy will be described.

\subsection{Angle encoding}\label{subsec:anglencoding}

\begin{figure}
    \centering
    \includegraphics[scale=1]{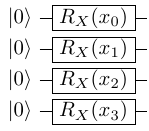}
    \caption{\label{fig:angle_encoding}Schematic example of the angle encoding method for a data-point with 4 features.}
\end{figure}

Frequently named as \textit{qubit encoding}, \textit{angle encoding} \cite{schuld_machine_2021, larose_robust_nodate} is one of the most simple data-encoding methods. It consists on encoding the features $x_k$ of the data point $x=(x_0, \dotsi, x_{M})$ in the amplitudes of the state-vector of each qubit:
\begin{equation}
    |x\rangle = \bigotimes_{k=0}^{M} \cos(x_k)|0\rangle + \sin(x_k)|1\rangle.
\end{equation}

where $M$ is the number of features in each data point and $M=n^2$ for square images. This encoding can be performed by applying $R_X(x_k)$ gates over each qubit (see Fig. \ref{fig:angle_encoding}), where the pixel $x_{i,j}$ is mapped to $x_k$ vector value with $k(i,j)=in+j$ index transformation. Due to the periodic nature of $R_X$ rotation gates, the classical features ought to be scaled to fit in $x_k \in [-\pi,\pi]$.

As the rotation gates acting on different qubits commute, angle encoding can be implemented in constant circuit depth, making it one of the most fit candidates to apply on NISQ devices. This implies $n^2$ queries to classical memory, so the time complexity would be $\mathcal{O}(n^2)$, where $n^2$ is the number of pixels (features). Regarding the number of qubits, this method requires one qubit per feature in the data-point, which can be reduced by a factor of two encoding an additional feature in the complex phase of each qubit, following the so-called \textit{dense angle encoding} strategy.

This improvement in the number of required qubits $n_q$, which is still linear with respect to the number of features, is notably surpassed by the exponential reduction achieved with \textit{amplitude encoding} \cite{schuld_machine_2021}. Nevertheless, in this type of encoding, the algorithm proposed by Mottonen et al.\cite{mottonen_transformation_2004} is usually employed for state preparation, which requires $\mathcal{O}(2^{n_q})$ multi-controlled gates. These gates do not commute generally, so the circuit depth of the implementation of Amplitude Encoding would be $\mathcal{O}(2^{n_q})= \mathcal{O}(2^{\log_2(n^2)}) =  \mathcal{O}(n^2)$. This depth requirements can be reduced by using ancillary qubits, and consequently loosing the advantage in the number of qubits; or encoding states with specific properties \cite{gonzalez-conde_efficient_2024,ashhabQuantumStatePreparation2022a, wangEfficientQuantumAlgorithm2009, weigoldExpandingDataEncoding2021}. \\
 
In general, angle encoding is one of the most employed encoding strategy in QML, in part for its simplicity, both in the implementation and for theoretical constructions. For this reason, angle encoding will be employed in this work, making use of a $n_q=16$-qubit system,  to encode images of resolution $n=4$. A way to implement the proposed algorithm for larger resolutions  will be discussed in subsequent stages. 

 \subsection{Equivariant variational circuit}\label{subsec:Equivariant variational}

\begin{figure}
\centering
    \begin{subfigure}[b]{0.45\columnwidth}
        \includegraphics[width=\columnwidth]{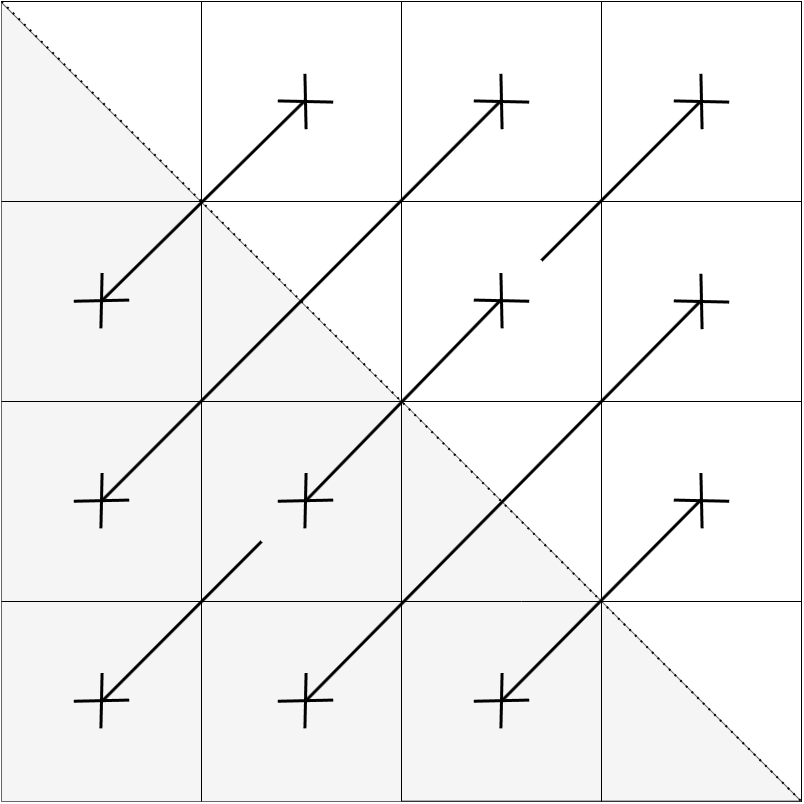}
        \caption{Transposition}
        \label{fig:trans}
    \end{subfigure}
    \hfill
    \begin{subfigure}[b]{0.45\columnwidth}
        \includegraphics[width=\columnwidth]{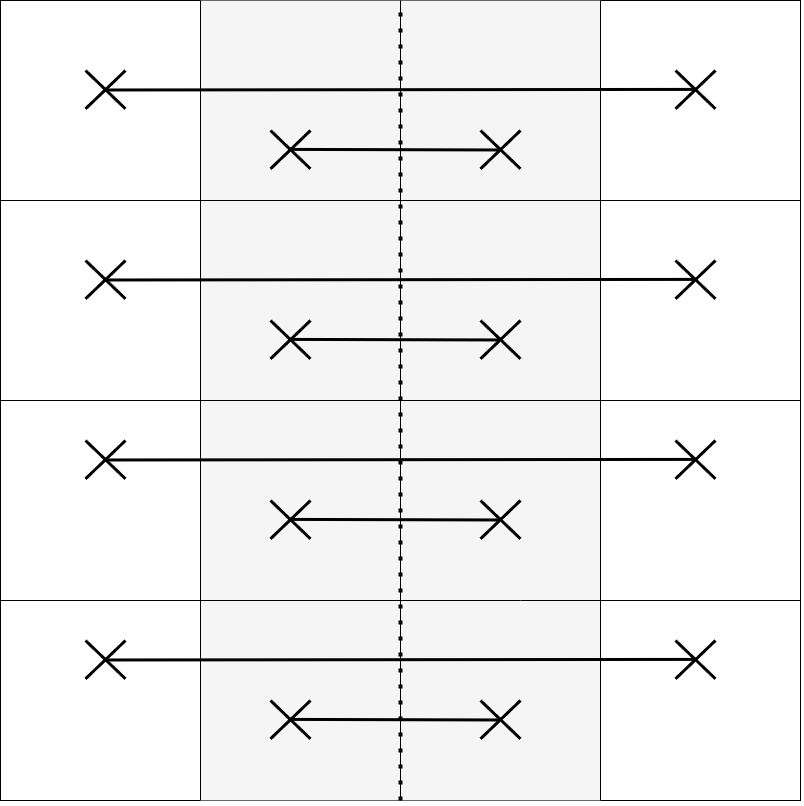}
        \caption{Column permutation}
        \label{fig:cols}
    \end{subfigure}
    \caption{\label{fig:trans+cols}Rotation operation decomposed into a (Fig. \ref{fig:trans}) transposition and (Fig. \ref{fig:cols}) column permutation using SWAP operations. Example for an image of n=4.}
\end{figure}

As a 4-fold rotation of an image can be performed by rearranging matrix elements according to the index transformation described in Eq. \ref{eq:rotation_index} and every matrix element is encoded in the quantum state of independent qubits, that symmetry operation can be represented in terms of hermitian \textit{SWAP} gates acting on the Hilbert space. The mentioned index transformation can be achieved by computing the transposed matrix of the original copy and reordering its columns (see Fig. \ref{fig:trans+cols}).

More specifically, for the chosen data encoding method, the quantum unitary that performs the action of the rotation symmetry element takes the following form:
\begin{eqnarray}\label{eq:Ug}
    U_g &=& \prod_{\substack{
    i=1,\dotsi,n\\
    j=1,\dotsi,\lfloor \frac{n}{2}\rfloor
    }}
    SWAP(q_{k(i,j)},q_{k(i,n-j-1)}) \nonumber \\
    &&
    \prod_{\substack{
    i=1,\dotsi,n\\
   j<i
    }}
    SWAP(q_{k(i,j)},q_{k(j,i)}).
\end{eqnarray}

The unitary representation $U_g$ of the symmetry element $g\in G$ can be used as a \textit{generator} for the representations of the rest of the symmetry elements in the group, operating by matrix multiplication with itself:
\begin{eqnarray}
    U_{g^2} = U_gU_g = U_g^2, \nonumber \\
    U_{g^3} = U_gU_{g^2} = U_g^3, \nonumber \\
    U_{g^4} = U_gU_{g^3} = U_g^4.
\end{eqnarray}
For this particular case, it is remarkable that $U_g^3$ represents a rotation in the opposite direction and, coping with unitary operations, $U_g^3=U_g^{-1}= U_g^\dag$. Moreover, it is straightforward that $U_g^4 =\mathbb{I}_{2^{n_q}}$ represents the identity operation. Thus, for this context, the symmetry group is the cyclic $C_4$ group and its unitary representation is $\mathcal{G}_U=\{U_g,U_g^2,U_g^\dag,\mathbb{I}\}$. Now it can be  verified that the equivariance condition described in Eq. \ref{eq:equiv_encoding} is satisfied for the chosen encoding unitary.

In order to build the variational blocks $W_l(\theta_l)$,  parameterized rotation gates will be used to rotate the quantum states of qubits in the Bloch's sphere and two-qubit CNOT gates to create entanglement between them. For convenience, the generators of the quantum gates will be simmetrized, which can be generated by \textit{Pauli operators}, since $[A,B]= 0 \Rightarrow [e^A,B]=0 $ :
\begin{equation}\label{eq:generator_pauli}
    R_{X}(\theta) \equiv e^{-i\theta\frac{X}{2}}.
\end{equation}
Now, it is possible to symmetrize Pauli operators using the twirling formula. If $X_k$ is defined as a Pauli gate $X$ applied on the qubit number $k$ (e.g. $X_0 = \mathbb{I}\otimes \mathbb{I} \otimes \dotsi \otimes X$), the following result is obtained:
 \begin{eqnarray}
    \mathcal{T}[X_k] &=& \frac{1}{4}(X_k+ U_g X_k U_g^\dag + U_g^2 X_k {U_g^\dag}^2 + U_g^3 X_k {U_g^\dag}^3) \nonumber \\
   & = & \frac{1}{4}(X_{k(i,j)}+ X_{k(n-j-1,i)}+ \nonumber \\
  &&  X_{k(n-i-1,n-j-1)} + X_{k(j,n-i-1)}).
 \end{eqnarray}

\begin{figure}
\centering
    \begin{subfigure}[b]{0.29\columnwidth}
        \includegraphics[width=\columnwidth]{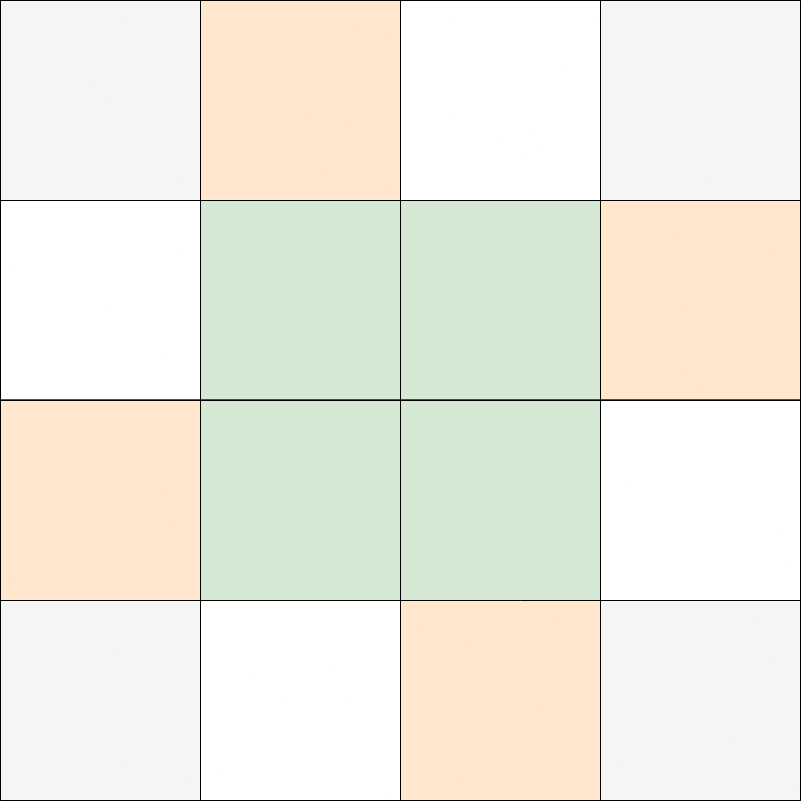}
        \caption{Orbits}
        \label{fig:orbits}
    \end{subfigure}
    \hfill
    \begin{subfigure}[b]{0.29\columnwidth}
        \includegraphics[width=\columnwidth]{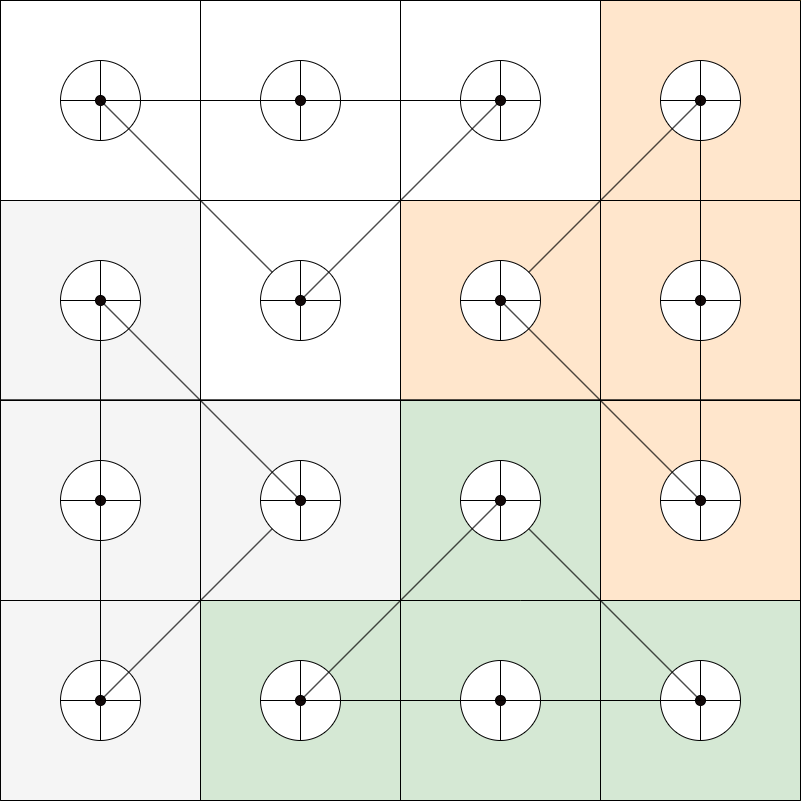}
        \caption{Intra-cell}
        \label{fig:intra}
    \end{subfigure}
    \hfill
    \begin{subfigure}[b]{0.29\columnwidth}
        \includegraphics[width=\columnwidth]{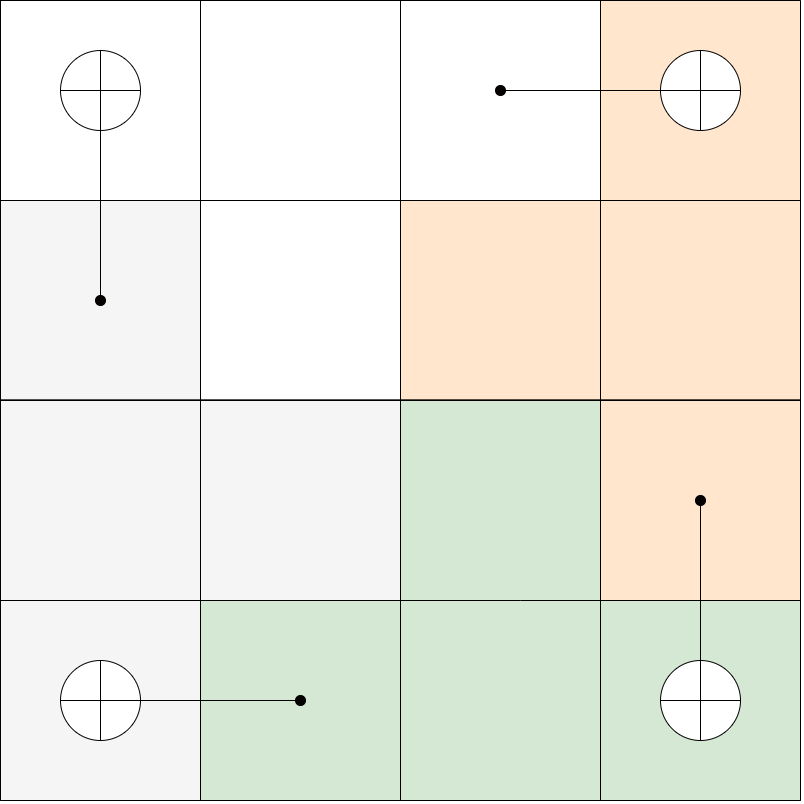}
        \caption{Inter-cell}
        \label{fig:inter}
    \end{subfigure}
    
    \caption{\ref{fig:orbits}) Different orbits of an image of resolution $n=4$ denoted by colors. \ref{fig:intra}) Different unit cells denoted by colors and the chosen circular connectivity of CNOT gates inside each unit cell, for an image of resolution $n=4$. \ref{fig:inter}) Different unit cells denoted by colors and the chosen circular connectivity of CNOT gates between unit cells, for an image of resolution $n=4$.}
    \label{fig:orbitcell}
\end{figure}

Let \textit{orbit} $\varphi$ denote the set of all different indices obtained under the action of all symmetry operations for a given pair of indices $(i,j)$ (see Fig. \ref{fig:orbits}):
\begin{eqnarray}
    \varphi(i,j)& =& \{(i,j), (n-j-1,i), \nonumber \\
    && (n-i-1,n-j-1), (j, n-i-1)\}.  
\end{eqnarray}

Then, the symmetrization of a Pauli gate is equal for every same Pauli gate applied on the qubits connected by the same orbit. This idea can be extended to the parameterized rotation gates generated by those Pauli gates since $[R_{X_k}(\theta), R_{X_{k'}}(\theta')] = 0$:
\begin{eqnarray}\label{eq:same_angle_rotations}
    &&e^{-i\frac{\theta}{8}(X_{\varphi_0(i,j)} + X_{\varphi_1(i,j)} + X_{\varphi_2(i,j)} + X_{\varphi_3(i,j)}) }  \\
    && \nonumber \\
    &=&R_{X_{\varphi_0(i,j)}}(\theta' )R_{X_{\varphi_1(i,j)}}(\theta')R_{X_{\varphi_2(i,j)}}(\theta')R_{X_{\varphi_3(i,j)}}(\theta')\nonumber.
\end{eqnarray}

where $\varphi_k(i,j)$ represents the component number $k$ of the orbit $\varphi(i,j)$. Consequently, the same symmetrization is obtained for rotation gates with the same rotation angle applied on qubits connected by the same orbit. This symmetrization is equivalent for the Y and Z rotation axes. The same logic can be followed for two-qubit gates, leading to this result: the same two-qubit gate must be applied over the orbits of both qubits, which must be in different orbits, to ensure the permutation condition of these gates with respect to the representation of the symmetry group. To illustrate this, as a reference to condensed matter physics, we will call \textit{unit cell} to the maximal set of indices that don't share orbits; this is, the maximal set of points that are not related by a symmetry operation (see Fig. \ref{fig:intra} and Fig. \ref{fig:inter}). 

Apart from the constraints concluded from the symmetrization of two-qubit gates, one has a certain freedom to choose their connectivity, i.e., on which qubit it can apply them. Indeed, the choice of the unit cell shown in Fig. \ref{fig:intra} and Fig. \ref{fig:inter} is not unique. A circular connectivity scheme will be selected, which entangles the qubits inside every unit cell. However, if different unit cells are not connected between them, it will produce four independent quantum circuits that would process just some few pixels of the original image. To avoid this scenario, some CNOTs can be added between unit cells and deal with a more extensive set of correlated qubits.

Once a variational block has been built out of symmetrized gates, the sequence of encoding and variational unitaries can be applied in an arbitrary number of layers to build an equivariant data re-uploading circuit. It is of great importance to make sure the chosen initial state of the quantum circuit is invariant to the action of the symmetry group, as mentioned in Sec. \ref{subsec:GQML}. In this case, it can verifed that $|0\rangle^{\otimes n_q}$ satisfies the invariant condition. Finally, an invariant observable is the last ingredient left to build a label-invariant model. A fit example for this problem can be the sum of Pauli Z gates applied on qubits connected by an orbit:
\begin{eqnarray}
    O &=& \frac{1}{4}(Z_0+Z_3+Z_{12}+Z_{15}), \nonumber\\
    f(x,\theta) &=& \langle\psi(x,\theta)|O|\psi(x,\theta) \rangle.
\end{eqnarray}

\subsection{Simulation results}\label{subsec:simulations}

An equivariant quantum circuit is trained to learn to classify tetromino images described in Sec. \ref{subsec:imagedata}. This circuit employs parameterized quantum gates symmetrized using the Twirling Formula, following the workflow detailed in Sec. \ref{subsec:Equivariant variational}. In consequence, a label-invariant model is obtained with respect to the action of 4-fold rotations. To benchmark the power of the model and to witness whether the geometric approach does provide any advantage in comparison to conventional variational circuits, its performance is compared to a widely used architecture called \textit{Basic Entangler Layers},  based on the work by Schuld, M. et al. \cite{schuld_circuit-centric_2020}. To make the comparison the fairest possible and since the latter architecture makes use of more optimizable parameters per layer, which can induce an improved approximation capacity, general rotation gates are used in the \textit{Equivariant} model, which can be decomposed into $3$ optimizable rotation gates, up to a negligible global phase \cite{nielsen_quantum_2010}:
\begin{equation}
    U = e^{i\alpha}R_Z(\beta)R_Y(\gamma)R_Z(\delta).
\end{equation}

It is true that these general rotation gates are more expressive than one-axis rotation gates employed in the Basic Entangler model, and this could be a possible source of a difference in performance. For this reason, a third model type is compared, which we will call \textit{Non-equivariant} model. Its quantum circuit utilizes the same architecture as the Equivariant model's but with randomly generated orbits to break the symmetrization of gates and, consequently, the equivariant condition. This way, it is ensured to capture if any improvement in the performance is due to the choice of gates or if it is, indeed, a consequence of the geometric approach.

In Tab. \ref{tab:params} the number of optimizable parameters per layer are shown for each model type. For the proposed equivariant architecture, the number of parameter increases linearly with the number of different orbits, which scales as $n_\varphi=n^2/4$ if $n$ is even, and $n_{\varphi}=\lceil\frac{n}{2}\rceil\lfloor\frac{n}{2}\rfloor+1$ if $n$ is odd . It is remarkable that this scaling does not stand true for the general case, since it is dependent on the gate symmetrization, determined by the chosen encoding method. For example, in the work of Chang, Su Yeon, et al. \cite{chang_approximately_2023}, and equivariant architecture was designed where it includes $6\lceil n_q/2\rceil+3\lceil n_q/4\rceil$ parameters per layer, and since amplitude encoding was employed, the scaling of the number of parameters per layer is logarithmic to the resolution of the image $n$.

\begin{table}
    \centering
    \begin{tabular}{|c|c|c|}
        \hline
        Model &  Parameters & Entangling gates\\
        \hline
         Equivariant & $3n_{\varphi}$ & $4(n_{\varphi}+1)$\\
        Non-equivariant & $3n_{\varphi}$ & $4(n_{\varphi}+1)$ \\
         Basic Entangler& $n^2$ & $n^2$  \\
         \hline
    \end{tabular}
    \caption{The number of optimizable parameters and entangling gates (CNOT) per re-uploading layer.  $n_{\varphi}$ corresponds to the number of different orbits in an image and $n$ is the resolution of the image. For an image of $n=4$, $n_{\varphi}=4$.}
    \label{tab:params}
\end{table}
\begin{figure}
    \centering
    \includegraphics[width=\columnwidth]{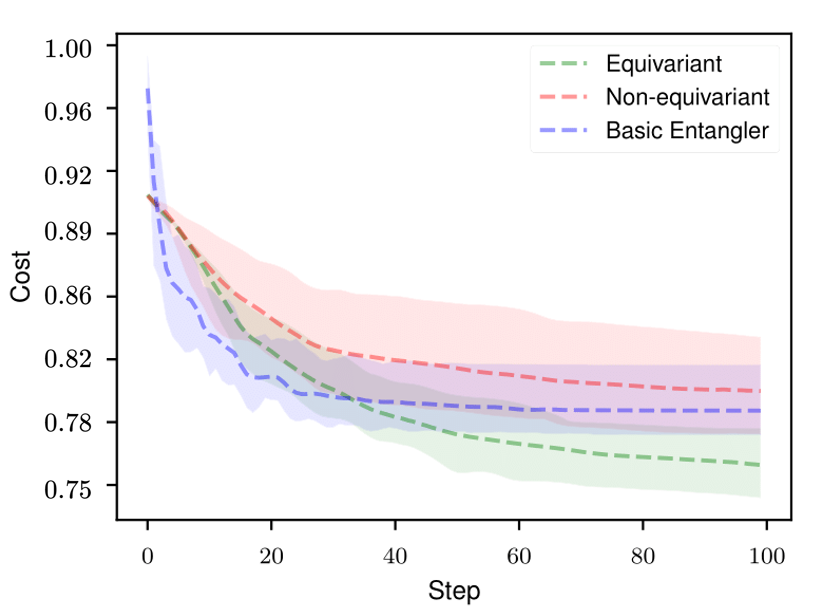}
    \caption{Loss landscape of three model types along the training process. Ten models with randomly initialized parameters were trained for each type of architecture (Equivariant, Non-equivariant and Basic Entangler). The shaded areas are spread between the maximum and minimum costs of each model type, and the dashed lines represent the mean cost of each model type. All models apply $n_l = 5$ re-uploading layers. These results correspond to the dataset of tetrominoes and a learning rate of $0.1$.}
    \label{fig: convergences}
\end{figure}

All three models are trained in a noiseless environment using the default simulator provided by \textit{Pennylane} \cite{bergholm_pennylane_2018}, and the parameter update is performed using \textit{Adam Optimizer} \cite{kingma_adam_2017}. In Fig. \ref{fig: convergences}, it can observed that the Equivariant model reaches the lowest values in the loss landscape for the same number of iterations. Even if the Basic Entangler performs a steeper descent in early steps, it reaches a plateau where the cost nearly stops decreasing. The other two models, however, keep reducing the cost until the last optimization step, although their descent tends to flatten in the last iterations. Comparing the gradients of the loss function for randomly sampled parameters, the Basic Entangler exhibits a variance two orders of magnitude smaller than that of the other two models. Similarly, the variance in loss values across different initializations is an order of magnitude smaller for the Basic Entangler model, indicating a more concentrated loss landscape compared to the other models (see Tab. \ref{tab:variances}). This aligns with the theoretical results of Ragone et al. (2023) \cite{ragone_unified_2023}: as symmetrization reduces the dimension of the associated Dynamical Lie Algebra (DLA), the loss landscape is expected to be less concentrated for the Equivariant and Non-Equivariant models.

\begin{table}[H]
    \centering 
    \resizebox{\columnwidth}{!}{
    \begin{tabular}{|c|c|c|c|c|}
    \hline
        &  $\widehat{E}_{\theta}[\mathcal{L}(x,\theta)]$ & $\Var_{\theta}{[\mathcal{L}(x,\theta)]}$ & $\widehat{E}_{\theta}[\frac{\partial \mathcal{L}(x,\theta)}{\partial \theta}]$ & $\Var_{\theta}{[\frac{\partial \mathcal{L}(x,\theta)}{\partial \theta}]}$\\
        \hline
        Equiv. & $0.999872$ &  \num{2.62e-3} &\num{-5.29e-05}& \num{8.25e-05}\\
        \hline
        Non-equiv. & $1.000313$ & \num{1.40e-3}& \num{6.30e-06}& \num{1.60e-05}\\
        \hline
         Basic Entangler. & $1.000109$ & \num{4.16e-4}& \num{1.39e-06}& \num{3.86e-07}\\
        \hline
    \end{tabular}}
    \caption{Expected values and variances of the loss function and its partial derivatives for randomly sampled parameters; comparison between Equivariant, Non-equivariant and Basic Entangler models respectively. 5000 parameter sets were uniformly sampled from $[0,2\pi)$ for each type of model.}
    \label{tab:variances}
\end{table}

\begin{figure}
    \centering
    \includegraphics[width=\columnwidth]{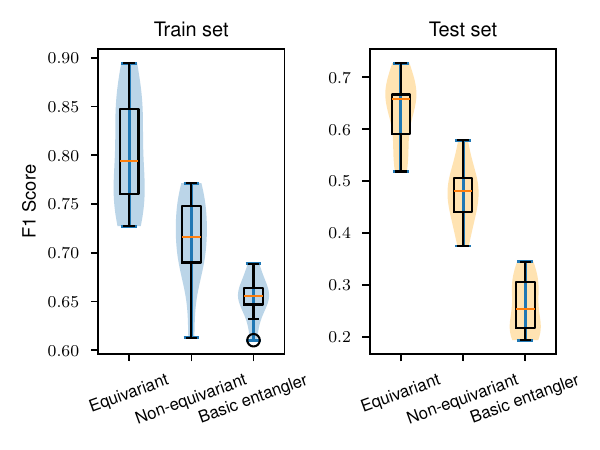}
    \caption{Violin plots of F1 score metrics of trained models. Ten different models of each type are trained with the same hyperparameters ($n_l=5$), and the performance is evaluated over the training and test set. These results correspond to the dataset of tetrominoes and a learning rate of $0.1$}
    \label{fig:violin}
\end{figure}

The loss function, defined as the mean quadratic distance between the circuit's predictions and the true labels, $\mathcal{L}(x,\hat{y}, \theta) = \frac{1}{N}\sum_i^N(f(x^{(i)},\theta)-\hat{y}^{(i)})^2$, does not capture accurately the performance of the models, and lower costs do not always result in better predictions. To evaluate and compare the predictive power of each model type, different metrics are used, such as the \textit{accuracy}, \textit{precision}, \textit{recall} and \textit{f1 score} \cite{tharwat_classification_2020}. In Fig. \ref{fig:violin}, the performance of various trained models is compared. It is remarkable how the equivariant models outperform the other two model types both in fitting the data and in generalization. 

Although the results shown in Fig.\ref{fig:violin} are favorable for the geometric solution, the performances are overall improvable, especially in their generalization capability. A classical procedure to handle this matter could be augmenting the data set size; nonetheless, the function classes the models can generate will be restricted by the number of data re-uploading layers applied in the quantum circuits. In consequence, the models will reach a performance ceiling that couldn't be improved even with more training steps. Performing additional data re-uploading layers will give the quantum circuit the capacity to generate higher-order trigonometrical polynomials of the input data features and, in conclusion, improve its approximation power. In Tab.\ref{tab:scores}, performance metrics for different numbers of data re-uploading layers are shown. 

\begin{table}
\centering
\resizebox{\columnwidth}{!}{%

\begin{tabular}{|l|c|c|c|c|c|}
\hline
Model Type & $n_l$ & F1 &  Accuracy &  Precision  & Recall  \\
\hline\hline
Equiv. & 2 & 0.59/0.47 & 0.66/0.52 & 0.72/0.57 & 0.51/0.40 \\
Non-equiv. & 2 & 0.53/0.50 & 0.61/0.53 & 0.65/0.58 & 0.45/0.44 \\
Basic Entangler & 2 & 0.51/0.47 & 0.55/0.47 & 0.54/0.50 & 0.49/0.45 \\
\hline
Equiv. & 4 & 0.78/0.63 & 0.81/0.64 & 0.85/0.68 & 0.73/0.59 \\
Non-equiv. & 4 & 0.55/0.41 & 0.62/0.48 & 0.66/0.52 & 0.47/0.34 \\
Basic Entangler & 4 & 0.47/0.37 & 0.55/0.46 & 0.55/0.48 & 0.41/0.31 \\
\hline

Equiv. & 6 & 0.79/0.67 & 0.82/0.70 & 0.87/0.77 & 0.73/0.61 \\
Non-equiv. & 6 & 0.63/0.45 & 0.68/0.54 & 0.71/0.60 & 0.57/0.37 \\
Basic Entangler & 6 & 0.67/0.46 & 0.73/0.52 & 0.80/0.56 & 0.59/0.41 \\
\hline

Equiv. & 8 & 0.69/0.48 & 0.75/0.58 & 0.85/0.72 & 0.58/0.36 \\
Non-equiv. & 8 & 0.63/0.50 & 0.68/0.57 & 0.75/0.69 & 0.54/0.41 \\
Basic Entangler & 8 & 0.66/0.47 & 0.72/0.52 & 0.79/0.58 & 0.56/0.41 \\
\hline
\textbf{Equiv.} & \textbf{10} & \textbf{0.98/0.93} & \textbf{0.98/0.93} & \textbf{1.00/1.00} & \textbf{0.96/0.88} \\
Non-equiv.& 10 & 0.73/0.54 & 0.78/0.60 & 0.82/0.65 & 0.67/0.47 \\
Basic Entangler & 10 & 0.87/0.73 & 0.87/0.74 & 0.89/0.81 & 0.85/0.67 \\
\hline
\end{tabular}
}

\caption{Performance metrics for each model type and per different number of data re-uploading layers. Each value in the table represents the mean performance metric of 5 different models, evaluated over the training set (left) and the test set (right). These results are for the dataset of tetrominoes.} \label{tab:scores}
\end{table}

\begin{figure}
    \centering
    \includegraphics[width=\columnwidth]{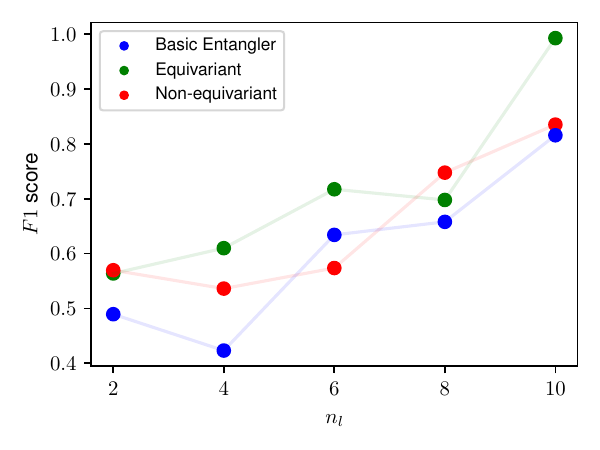}
    \caption{Mean $F1$ scores over the test set of the three model types (10 modes per type) in function of the number of data re-uploading layers $n_l.$}
    \label{fig:reuploadings}
\end{figure}

The results in Tab.\ref{tab:scores} show, on one hand, that, in general, the Equivariant model achieves better performances both in the training and in the test sets. Moreover, it is noticeable how increasing the number of data re-uploading layers tends to improve the approximation ability of all three models; this is well illustrated in Fig.\ref{fig:reuploadings}. 

\subsection{Scaling the problem size}\label{subsec:scalling}
The data set employed in the previous sections represents a toy case created for the comparison of the capabilities of the geometric approach with respect to the traditional variational quantum models. The resolution of images in real computer vision problems, however, is far from the few pixels the current NISQ-era devices can handle due to their limited circuit size. A way to address this issue is to use a another encoding strategy, different from the one proposed in this work, that optimizes the number of features encoded per qubit. Another way to cope with this matter, rather than waiting to larger-scale quantum devices to be built and implement the proposed architecture for larger images, is to reduce the dimensionality of the input.

\begin{figure}
    \centering
    \includegraphics[width=\columnwidth]{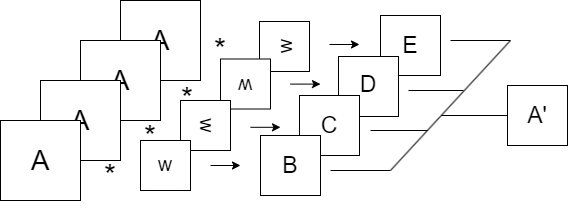}
    \caption{Rotation equivariant convolution operation.}
    \label{fig:equiv_conv}
\end{figure}

There are several feature reduction techniques such as PCA, as the main lineal transformation method, or autoencoders among the non-linear transformations \cite{van_der_maaten_dimensionality_2007}. In this work, we propose to use trainable convolutional layers to extract features in the latent space, preserving the spatial arrangement of the features in the original image. Nevertheless, if the model is wanted to be label-invariant, the classical convolution layers must be equivariant to the operations in the symmetry group. A two-dimensional discrete convolution between an image $x$ and a filter $w$ can be described by the following equation, assuming square images and filters \cite{gonzalez_digital_2018}:
\begin{equation}
   ( x \ast w)[i,j] = \sum_{s=-r}^{r}\sum_{t=-r}^r x[s,t]w[i-s,j-t],
\end{equation}
where $r = (n_f-1)/2$ for odd $n_f$ and $r=n_f/2$ for even $n_f$, being $n_f$ the size of the filter. This expression can be extended for multichannel images, RGB images for example,  by adding an extra dimension to the convolution filters, so the convolution filter acts over $n_c$ channels of the input tensor. However, in general, this operation is not equivariant to image rotations described by Eq. \ref{eq:rotation_index}. For this reason, following the approach of Cohen, T., \& Welling, M. \cite{cohen_group_2016}, a rotation-equivariant convolution operation is defined by computing a convolution between an image and all rotated versions of the original filter, and averaging all resultant products into a single output (see Fig. \ref{fig:equiv_conv}).

These equivariant convolutional layers can be stacked arbitrarily in $n_c'$ channels, each one of them with a different filter so that they learn to focus on specific characteristics of the images. This way, if $n_x$ is the size (number of pixels) of the rows of the square image, $n_w$ the size of the filter, $s$ the stride and $p$ the padding, the size of the convoluted image will be \cite{dumoulin_guide_2018}:

\begin{equation}
    n'_x  = \frac{n_x+2p-n_w}{s}+1
\end{equation}

The stride is the number of pixels the convolutional filter moves across the input image each time it slides and the padding is the addition of pixels (often zeros) around the input image's border. For this case, $p=0$ and $s=1$ will be taken, so $n'_x = n_x-n_w+1$. Taking the data set as a tensor of $N$ data-points allows us to apply the equivariant convolution operation over the whole batch, performing the following dimensional reduction:
\begin{eqnarray}
    \textrm{Input tensor:}&\; & (N,n_x,n_x,n_c) \nonumber\\
    \textrm{Filter:} &\; &(n_w,n_w,n_c,n'_c) \nonumber\\
    \textrm{Output tensor} &\;& (N,n'_x, n'_x, n'_c).
\end{eqnarray}

Accordingly, the number of channels of each convolutional layer and the size of the filters can be tuned to reduce the number of features and adapt the input to the size of the available quantum circuit.  In this case, to connect the classical part to the equivariant quantum classifier described in the previous sections, the original images are reduced to $4\times4$, single channel square images. It is postulated that while the number of qubits keeps increasing and the connectivity between them condensing, the size of the images that can be embedded directly into the quantum classifier will increase while decreasing the dependency on the classical convolutional part. 

Therefore, the final model is constituted by a set of classical equivariant convolution layers, each one of them with optimizable parameters, that return an output in a reduced dimensional latent space, as an input for the rotation-invariant quantum model. The prediction of the model is evaluated in a mean quadratic distance loss function and the parameter update, both for the classical and quantum part, is done using Adam Optimizer to minimize the loss function, in the same fashion as in Sec. \ref{subsec:simulations}. After each convolution layer, a \textit{ReLU} activation function is applied in order to make the classical part learn nonlinearities \cite{DUBEY202292}. Average pooling layers are applied between convolutions to reduce the dimensionality of the tensors. It can be verified that, unlike other types of pooling operations, average pooling is equivariant to image rotations as it is equivalent to performing a convolution with equal and normalized filter elements.

This algorithm is tested on three public datasets. The first one, \textit{Fashion MNIST}\footnote{\url{https://huggingface.co/datasets/fashion_mnist}}, contains $28\times28$ grayscale images. For binary classification, labels $0$ (T-shirts) and $1$ (trousers) have been taken. The second dataset, \textit{11k Hands}\footnote{\url{https://sites.google.com/view/11khands}} \cite{afifi201911kHands}, contains $256\times256$ RGB images of hands and the model is required to classify between left and right hands. It has been chosen as an illustrative example of a rotational symmetric problem where reflection symmetry is not present. Both previous datasets have been augmented by adding rotated versions of original images. The last dataset, \emph{Resisc45}\footnote{\url{https://huggingface.co/datasets/timm/resisc45}} \cite{chengRemoteSensingImage2017}, contains satellital $256\times256$ RGB images, and the model is required to classify between \textit{islands} and \textit{lakes}. All datasets are split with a test/train ratio of $1/3$. For reproducibility, the chosen hyperparameters for training the extended quantum models over each dataset are shown in Tab. \ref{tab:hyperparams}.

\begin{table}
\centering
\resizebox{\columnwidth}{!}{%

\begin{tabular}{|l|c|c|c|c|c|c|c|}
\hline
Dataset & $n_w$ & $n_c$ &  $n_p$ &  $n_l$  & Dataset Size & Max. it. & L.r.  \\
\hline\hline
MNIST & 11,11,3,3 & 10,10,10,1 & - & 6 & 200 & 100 & 0.001 \\
11kHands & 3,2,2,4,7 & 8,64,64,6,1 & 2,2,2,2,2 & 6 & 100 & 1000 & 0.001 \\
Resisc45 & 3,2,2,4,7 & 32,32,32,32,1 & 2,2,2,2,2 & 4 & 500 & 100 & 0.01\\
\hline

\end{tabular}
}

\caption{Chosen hyperparameters for training the extended quantum models for each dataset. $n_w$ represents the dimension of each convolution filter, $n_c$ the number of channels in every convolutional layer, $n_p$ the dimension of each pooling filter, $n_l$ the number of re-uploading layers, \textit{Max. it.} is the number of training epochs and \textit{L.r.} is the learning rate of the optimizer.} \label{tab:hyperparams}
\end{table}

Keeping the classical convolutional layers equal, different models with \textit{Equivariant}, \textit{Non-equivariant} and \textit{Basic-Entangler} quantum circuits are trained over each dataset to compare the performances of the architectures. In Tab. \ref{tab:extended_scores} the performance metrics of each extended model are shown, for different datasets. It can be seen that the \textit{Equivariant} models tend to outperform the other two architectures, pointing that the geometric approach does indeed bring a classification advantage where image rotation label symmetry is present. This is not true in the case of MNIST dataset, where \textit{Equivariant} and \textit{Non-equivariant} models show similar metrics. It is important to consider that classical convolution layers contribute significantly to the learning process, and there can be cases where the classical layers process the information enough to leave a trivial classification problem to the quantum part, vanishing any advantage coming from the geometric approach.

\begin{table}
\centering
\resizebox{\columnwidth}{!}{%

\begin{tabular}{|c|c|c|c|c|c|}
\hline
Dataset & Model Type  & F1 &  Accuracy &  Precision  & Recall  \\
\hline\hline

MNIST &Equiv. & 0.91/0.89 & 0.89/0.86 & 0.89/0.87 & 0.88/0.86 \\
MNIST &Non-equiv. & 0.92/0.88 & 0.90/0.86 & 0.90/0.87 & 0.90/0.96 \\
MNIST &Basic Entangler & 0.83/0.81 & 0.77/0.75 & 0.76/0.76 & 0.77/0.75  \\
\hline
11kHands & Equiv. &  0.84/0.40 & 0.85/0.57 & 0.83/0.43 & 0.86/0.57 \\
11kHands &Non-equiv. & 0.53/0.24 & 0.71/0.60 & 0.60/0.38 & 0.71/0.60 \\
11kHands & Basic Entangler & 0.57/0.28 & 0.78/0.62 & 0.65/0.36 & 0.78/0.62 \\
\hline

Resisc45 & Equiv. &  0.64/0.64 & 0.78/0.80 & 0.59/0.88 & 0.78/0.80 \\
Resisc45 &Non-equiv. & 0.52/0.52 & 0.70/0.71 & 0.72/0.86 & 0.70/0.71 \\
Resisc45 & Basic Entangler & 0.38/0.44 & 0.64/0.68 & 0.53/0.88 & 0.64/0.68 \\
\hline

\end{tabular}
}

\caption{Performance metrics for each extended model type trained over different datasets. Each value in the table represents the mean performance metric, evaluated over the training set (left) and the test set (right). These metrics can be improved with additional computational resources, such as a deeper convolutional part, more re-uploading layers or a larger data set.} \label{tab:extended_scores}
\end{table}

\begin{figure}
    \centering
    \includegraphics[width=\columnwidth]{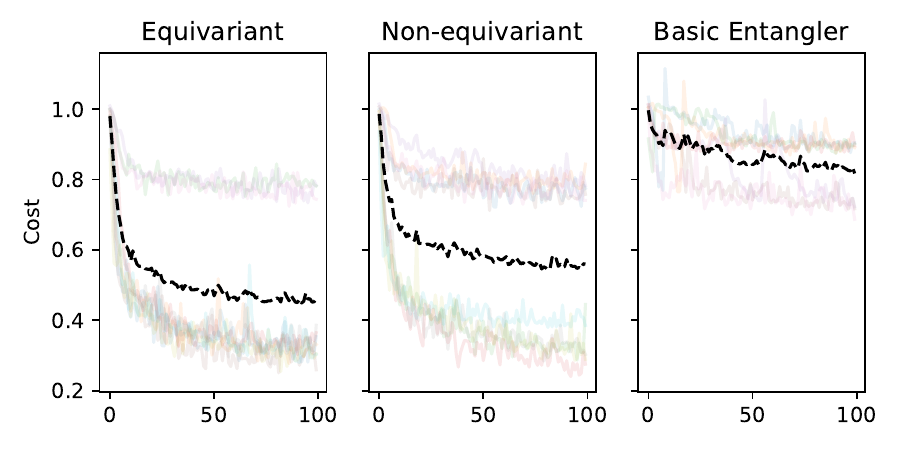}
    \caption{Cost history during the training of models. The black curve represents the mean cost in each step. Depending on the initialization of the parameters, the models get stuck in local minima. These results are for Resisc45 dataset, but the same phenomenon is observed for other datasets. 10 different initializations are shown for each model type.}
    \label{fig:extended_cost}
\end{figure}

Finally, during the model training process, a loss landscape characterized by numerous local minima \cite{anschuetz_quantum_2022} was observed (see Fig. \ref{fig:extended_cost}). From the study of the distribution of minima, we can extract some conclusions about the loss landscapes of each type of model. First, Tab. \ref{tab:minima} shows that both the Equivariant and Non-equivariant models reach lower minima and they have a higher density in the minima below the mean loss value, comparing with the Basic Entangler model. According to Larocca M., et. al. \cite{larocca_theory_2023}, the required number of parameters to reach overparametrization is directly related to the dimension of the associated Dynamical Lie Algebra. In our case, all three models contain approximately the same number of parameters, but the dimension of the associated DLA is lower in the first two models. Therefore, the limit of overparametrization is closer for these two, and the trainability is expected to be improved compared to the Basic Entangler model. The observations agree with this principle. Appart from that, it can be seen how the lowest minima reached in every dataset correspond to the Equivariant model indicating that the introduced geometric bias helps the parameterized quantum circuit to produce better models. The study of the effect of initial points in the optimization process is considered of great interest and delegated to future work.

\begin{table}[h]
    \centering
    \resizebox{\columnwidth}{!}{%
    \begin{tabular}{|c|c|c|c|c|}
        \hline
        Dataset & Model Type  & Mean Loss  &  Min Loss &  \% Below Mean \\
        \hline\hline
        Tetrominoes &Equiv & $0.83$ & $0.76$ & 51  \\
        Tetrominoes &Non-equiv &$ 0.91$&$0.85$& $51$  \\
        Tetrominoes &Basic Entangler &$0.97$& $0.96$ & $43$   \\
        \hline
        MNIST &Equiv &$0.41$ & $0.04$ & $62$  \\
        MNIST &Non-equiv & $0.34$ &$0.1$& $62$ \\
        MNIST &Basic Entangler &$0.7$& $0.27$ & $48$  \\
        \hline
        11kHands & Equiv & $0.6$ & $0.24$&$ 65$  \\
        11kHands &Non-equiv & $0.71$& $0.34$&$70$  \\
        11kHands & Basic Entangler & $0.77$& $0.45$ & $44$  \\
        \hline
        Resisc45 & Equiv & $0.45$ &$0.25$& $70$  \\
        Resisc45 & Non-equiv & $0.55$& $0.27 $& $50$  \\
        Resisc45 & Basic Entangler &$0.81$ & $0.68$ & $40$ \\
        \hline
    \end{tabular}
    }
    \caption{Mean and minimum values of the loss function after the training of various randomly initialized parameters. The last column represents the percentage of loss observations below the mean value. Respectively, in order of appearance in the table, $35$, $21$, $9$ and $10$ random initializations were tried per model type in each dataset.}
    \label{tab:minima}
\end{table}

\section{Conclusion}\label{sec:conclusion}

This work has shown a methodology to build a label-invariant VQC for image classification with $C_4$ rotation label-symmetry. This has been achieved by symmetrizing the quantum gates using the Twirling formula and choosing the right initial state and observable. The model has been benchmarked in a synthetic data set, and compared against two quantum models with different architectures. The first one is a standard architecture, widely used in the literature; the second uses the same gates, initial state, and observable as the label-invariant model but with randomized orbits, breaking any symmetry-preserving relation.

Experimental results in simulators have shown that the equivariant quantum circuit obtains better performances than the other two models, coinciding with the expected results from the literature of GQML. It reaches lower minima than the two other models and the performance metrics, both in the training and test sets, are higher on average. The difference in the performance between the \textit{Equivariant} and \textit{Non-equivariant} models reveals that the improvement in classification comes from the geometric approach since the same gates, observable, and number of parameters have been used, and the only change is the qubits they are acting on. Alternatively, the importance of data re-uploading has been observed in the approximation capabilities of every model, increasing the performance metrics with the number of re-uploading layers.

The quantum models were designed for a synthetic toy data set of phew grayscale pixels, considering the reduced number of available qubits and low connectivity in NISQ devices, and the computational cost of simulating them. An equivariant convolution operation has been proposed to implement the quantum circuits for higher-resolution images with color channels while preserving the label-invariance of the model. This way, the weights of convolution filters have been trained to learn the most essential features of the image, which are introduced as input to the quantum circuit. This hybrid algorithm has been tested on three public datasets and the three quantum architectures have been compared, obtaining again the best results for the geometric model. Numerical results have shown that the reduction of the associated Dynamical Lie Algebra, improves the trainability of the model by alleviating the problems of loss concentration (Barren Plateaus) and local minima, coinciding with theoretical results in literature. Gate symmetrization is a powerful tool for reducing the DLA while introducing inducted bias of the problem, improving its performance.

Taking into account that all the experiments have been made in an ideal, noise-free simulator, we consider it of great interest to execute the models in real quantum hardware and observe their behavior with noise. In addition, the chosen data encoding method embeds a single grayscale value per qubit. The most straightforward solution to embed color channels could be to execute parallel equivariant circuits, one for each channel, entangled between them. Another approach could be to use the imaginary phase of qubits and map every point of Bloch's sphere to a two-dimensional colormap using dense angle encoding. We delegate to future investigations to research qubit-efficient data encoding methods that consider color channels and are equivariant to image rotations. The choice of the encoding will determine the unitary representation of the symmetry group and, hence, the symmetrization of gates. Extending the model for multi-class classification could be another task for future work, choosing the right observable that returns multiple labels and preserves the symmetry relations of the data.

Finally, we believe variational quantum algorithms have great potential as an early application of NISQ devices, although they still have significant limitations \cite{cerezo_does_2023}. All in all, lots of work is to be done, and the geometric framework seems a powerful tool to design new architectures for machine learning models, even beyond quantum algorithms. We consider this work contributes in this direction, first, providing experimental evidence of the advantages of adding geometric inductive bias to the design of quantum circuits. Additionally, the employed methodology may serve as a blueprint for other research teams to apply GQML to image classification problems and push its state-of-the-art. We encourage researchers to use these results as a reference to benchmark and develop new applications. To conclude, we identify future research lines that extend this work, connecting theoretical concepts with experimental work (and vice versa), and interesting results are expected from them. We hope this work serves as a step in the path toward the utility of variational quantum algorithms.

\bigskip 
 
{\bf Acknowledgements.-} We acknowledge Ikerlan, DIPC, Ikerbasque, Basque Government, Diputaci\'on de Gipuzkoa and EIC for support, as well as discussions with the teams from Multiverse Computing and Ikerlan. This article has been supported by the  Government of the Basque Country through the research grants ELKARTEK KUBIT - Grant number KK-2024/00105 (Kuantikaren Berrikuntzarako Ikasketa Teknologikoa),  and the Gipuzkoa Provincial Council through the research grant QIA, grant number 2023-QUAN-000031-01 (Konputazio Kuantikoa Adimen Artifizialeko algoritmoetan: Konputazio klasikotik Quantum Machine Learning-era)

\newpage
\bibliography{references.bib}

\end{document}